\documentclass[english,aps,preprint,prl,twocolumn]{revtex4-1}
\usepackage[T1]{fontenc}
\usepackage[latin9]{inputenc}
\usepackage{amsmath}
\usepackage{esint}

\makeatletter

\@ifundefined{textcolor}{}
{%
 \definecolor{BLACK}{gray}{0}
 \definecolor{WHITE}{gray}{1}
 \definecolor{RED}{rgb}{1,0,0}
 \definecolor{GREEN}{rgb}{0,1,0}
 \definecolor{BLUE}{rgb}{0,0,1}
 \definecolor{CYAN}{cmyk}{1,0,0,0}
 \definecolor{MAGENTA}{cmyk}{0,1,0,0}
 \definecolor{YELLOW}{cmyk}{0,0,1,0}
}

\usepackage{babel}

\makeatother

\usepackage{babel}
\begin{document}

\preprint{\bibliographystyle{revtex}}


\title{Non-Equilibrium is Different}

\author{T. R. Kirkpatrick and J. R. Dorfman}

\affiliation{Institute for Physical Science and Technology, University of Maryland,
College Park, MD 20742}
\date{\today}

\begin{abstract}
Non-equilibrium and equilibrium fluid systems differ due
to the existence of long-range correlations in non-equilibrium that are not
present in equilibrium, except at critical points. Here we examine
fluctuations of the temperature, of the pressure tensor, and of the
heat current in a fluid maintained in a non-equilibrium stationary
state (NESS) with a fixed temperature gradient, a system where the non-equilibrium correlations are especially long ranged. For this particular NESS our results show that
(1) The mean-squared fluctuations in non-equilibrium differ
markedly in their system size scaling compared to their equilibrium
counterparts and (2) There are large, nonlocal,
correlations of the normal stress in this NESS. These terms provide
important corrections to the fluctuating normal stress in linearized
Landau-Lifshitz fluctuating hydrodynamics. 
\end{abstract}
\maketitle
In recent years fluctuations in fluids maintained in non-equilibrium
steady states (NESS), and in non-equilibrium fluids in general, have
attracted a large amount of attention. Of particular interest are
the studies of Evans, Cohen and Morris (ECM) \cite{Evans_Cohen_Morris_1993},
Evans and Searles \cite{Evans_Searles_1994}, and of Gallavotti and
Cohen \cite{Gallavotti_Cohen_1995}. Later related work was done by
Jarzynski \cite{Jarzynski_1997} and by Crooks \cite{Crooks_1999}.
Particularly in the earlier work, a focus was on non-equilibrium currents
and entropy production. For example, if $P_{xy\tau}$ is the time
average of the microscopic stress tensor, $P_{xy}$, over a time interval
$\tau$ then in a NESS with a steady shear rate, $\gamma$, ECM studied
the probability distribution $\mathcal{P}(P_{xy\tau})$ in a $N-$particle
system. Using an analysis based upon their computer simulations, they
showed that, 
\begin{equation}
\frac{\mathcal{P}(P_{xy\tau})}{\mathcal{P}(-P_{xy\tau})}=\exp\left\{ N\gamma\int_{0}^{\tau}d\tau P_{xy}[\varGamma(\tau)]\right\} 
\end{equation}
On the RHS the phase space time-dependence $\varGamma(\tau)$ is restricted
such that only the dynamical states $i$ with a given value of $P_{xy\tau}=\int_{0}^{\tau}d\tau P_{xy}[\varGamma_{i}(\tau)]$
are taken into account. They also numerically determined $\mathcal{P}(P_{xy\tau})$
and found it had a sharp peak around it's average value and that although
the bulk of the distribution was consistent with positive entropy
production, there was a tail part consistent with negative entropy
production. The system size or $N$-dependence of the width of the
distribution, $\mathcal{P}(P_{xy\tau})$ is an interesting open problem.

Prior to this work, it was established in the $1960's$ through the
$1980's$ that correlations are very different in non-equilibrium
fluid systems than they are equilibrium systems \cite{Dorfman_Cohen_1965,Peierls_1979,Dorfman_Kirkpatrick_Sengers_1994}. For example, in the early $1980's$
it was predicted \cite{Kirkpatrick_Cohen_Dorfman_1982A}
that the temperature and density correlations in a fluid in a NESS
with a temperature gradient were extraordinarily long-ranged (in a
sense growing with system size). This prediction was subsequently
confirmed \cite{Law_et_al_1990,DeZarate_Sengers_2006} with great
precision in small angle light scattering experiments. For very small
angle scattering, the scattering was found to be larger than the equilibrium
scattering by a factor of $10^{5}$. All of this implies that the
statistics of fluctuations in non-equilibrium fluids will in general
be very different than those for fluctuations in the same fluid in
an equilibrium state. For reviews see \cite{Dorfman_Kirkpatrick_Sengers_1994,Schmittmann_Zia_1995,DeZarate_Sengers_2006,Derrida_2007}

Here we will expand on this point by examining the system size dependence of not only the temperature fluctuations and distribution, but also the fluctuations and distributions of the pressure tensor,
and heat current for a fluid in
a NESS with a temperature gradient. This particular NESS (see also point 6 in the discussion) is unique because the correlations are so strong that they extend over the entire system size, see Eq.(\ref{eq:delbarns}). This is very different then the correlations in other NESS, which typically decay as a power law. We find that that temperature
and pressure or normal stress tensor fluctuations in a fluid with a temperature gradient are so large that many
of the assumptions that are commonly made about non-equilibrium fluids
when one uses equilibrium-like methods, such as maximum entropy formalisms,
power series expansions, etc. to describe their properties, are
called into question and may not be justified. We show that relative
root mean square fluctuations of some thermodynamic and transport quantities
do not in general scale with particle number, $N,$ as $1/N^{1/2}$ as do their
equilibrium counterparts. We also describe the
distributions of these fluctuations, and point out their anomalous
properties as well. 

The new results in this paper are: (1) We show that the previously computed temperature fluctuations for a fluid in a temperature gradient scale as $1/N^{(d-2)/2}$ and not as $1/N^{1/2}$, (2) We compute for the first time the pressure or stress fluctuations for a fluid in a temperature gradient. These fluctuations are shown to scale as $1/N^{(d-2)}$, (3) We compute for the first time the heat current fluctuations for a fluid in a temperature gradient. These fluctuations do not exhibit anamolous scaling, (4) We discuss the implications of these fluctuation results for the distributions of temperature, pressure, and heat current, (5) We argue that the linearized about the NESS Landau and Lifshitz fluctuating hydrodyanmics are not consistent with the long range nature of the normal stress fluctuations.

We start with the well-known expression for the small wavenumber behavior
of the temperature fluctuations \cite{Kirkpatrick_Cohen_Dorfman_1982A,DeZarate_Sengers_2006},
\begin{eqnarray}
\left\langle \left|\delta T({\bf k})\right|^{2}\right\rangle _{NESS} & = & \frac{k_{B}T}{\rho D_{T}(\nu+D_{T})}\frac{(k_{\parallel}\nabla T)^{2}}{k^{6}}.\label{eq:delTk}
\end{eqnarray}
Here $\rho$, $\nu$ and $D_{T}$ are the mass density, the kinematic
viscosity, and thermal diffusivity of the fluid. This result is valid as long as $k>1/L$, with $L$ the system size. Here we have assumed that the temperature
gradient is in, say, the $z$-direction so that $k_{\parallel}=\sqrt{k_x^2+k_y^2}$ is
the magnitude of the wavenumber parallel to the confining walls. We note that this correlation function is long ranged
as indicated by its $k^{-4}$ behavior at small wave numbers, while
as the equilibrium temperature fluctuations are localized in space
with no singular behavior of the corresponding Fourier transforms
at small wave numbers. In order to determine the spatial correlation
of the temperature fluctuations one must invert the Fourier transform
to obtain the correlation function $C_{T,NESS}({\bf r})$
\cite{DeZarate_Sengers_2006}. Using Eq. (\ref{eq:delTk}) we find
that the spatial average of the temperature correlations has a strikingly
different size dependence than the corresponding equilibrium temperature
correlation. That is 
\begin{eqnarray}
 \bar{\Delta}_{T,NESS}\propto\frac{k_{B}T}{\rho D_{T}(\nu+D_{T})}L^{4-d}(\nabla T)^{2},\label{eq:delbarns}
\end{eqnarray}
 where 
\begin{equation}
\bar{\Delta}_{T,NESS}=\frac{1}{L^{d}}\int d{\bf r}C_{T,NESS}({\bf r}),\label{eq:delbardef}
\end{equation}
 where $L$ is the characteristic system size. In general we will
suppress numerical factors in spatially averaged quantities. The $N$
dependence of this spatial average is easily found to be of order
$N^{(4-d)/d}$while the equilibrium quantity would be 
\begin{equation}
\bar{\Delta}_{T,eq}=\frac{k_{B}T^{2}}{L^{d}\rho c_{v}},\label{eq:delbareq}
\end{equation}
where $c_{v}$ is the specific heat per mass at constant volume. The $N$-dependence
quoted above for Eq.(\ref{eq:delbarns}) is for fixed $\nabla T$.
Physically, and experimentally, it may be more reasonable to use $\nabla T=\frac{\Delta T}{L}$
with $\Delta T$ the temperature distance between the fluid confining
plates in the direction of the temperature gradient and fix $\Delta T$.
In this case the scaling of Eq.(\ref{eq:delbarns}) is $\bar{\Delta}_{T,NESS}\propto1/L^{d-2}\propto1/N^{(d-2)/d}$.
In either case $\bar{\Delta}_{T,NESS}$ is large compared to $\bar{\Delta}_{T,eq}$
in the large system size limit. Below we generally fix $\Delta T$,
to examine the system size dependence. The theoretical results appear
even more anomalous if the temperature gradient is fixed.

For three dimensional systems a natural length, $l$, that occurs
is
\begin{equation}
l=\frac{k_{B}T}{\rho D_{T}(\nu+D_{T})}.\label{eq:ldef}
\end{equation}
For water at STP, $l\approx3\times10^{-9}cm$. The ratio of the non-equilibrium
to equilibrium temperature fluctuations is 
\begin{eqnarray}
 \frac{\bar{\Delta}_{T,NESS}}{\bar{\Delta}_{T,eq}}=\left(\frac{\rho c_{v}\ell}{nk_{B}\sigma}\right)n\sigma^{3}\left(\frac{\Delta T}{T}\right)^{2}\left(\frac{L}{\sigma}\right)^{2}.\label{eq:epsilonT}
\end{eqnarray}
with $\sigma$ a molecular diameter. This ratio is generally large
if $L\gg\sigma$. Taking the reduced density to be unity, for water
this requires, roughly, 
\begin{equation}
\frac{L}{\sigma}>\epsilon =\frac{T}{\Delta T}\left(\frac{nk_{B}\sigma}{\rho c_{v}\ell}\right)^{1/2}\label{eq:Lovrsig}
\end{equation}
or, $L/\sigma>10$ for typical experiments\cite{Law_et_al_1990,DeZarate_Sengers_2006}
where $\Delta T/T<1/5.$ This sets the scale of the system size where
non-equilibrium fluctuations become dominant. If we assume that the
temperature fluctuations in the NESS have a Gaussian distribution,
we find, again for three dimensions, that
\begin{equation}
\mathcal{P}_{NESS}[\delta T]\sim\exp[-\frac{L}{2\ell}\frac{(\delta T)^{2}}{(\Delta T)^{2}}]\label{eq:probdelT}
\end{equation}
 The corresponding probability distribution for equilibrium temperature
fluctuations is 
\begin{equation}
\mathcal{P}_{eq}[\delta T]\sim\exp[-L^{3}\frac{\rho c_{v}(\delta T)^{2}}{2k_{B}T^{2}}].\label{eq:eqprbdelT}
\end{equation}
From a comparison of these two distributions one can see that the
non-equilibrium distribution is dominant whenever the inequality given
by Eq.(\ref{eq:Lovrsig}) is satisfied. From Eq.(\ref{eq:probdelT})
we see that in non-equilibrium the temperature fluctuations scale
as $\delta T\sim1/L^{1/2}\sim1/N^{1/6}$ in three-dimensional systems
(compared to $1/N^{1/2}$ in equilibrium systems).

Next we consider pressure or normal stress fluctuations
in this NESS. It is important to note that the average pressure or normal stress
at some point in the fluid has a non-equilibrium term that is proportional to the square
of the temperature gradient. This follows from direct calculations,
but it can be understood by symmetry arguments as well, since the
pressure, for example, is a scalar quantity. The result for the most important non-equilibrium
part of the (spatial and thermal) average pressure in three-dimensions
is singular in the limit of large systems and is found to be \cite{Kirkpatrick_DeZarate_Sengers_2013,Kirkpatrick_DeZarate_Sengers_2014}

\begin{equation}
P_{NESS}\propto A\ell L\left(\nabla T\right)^{2}=A\ell\frac{(\Delta T)^{2}}{L},\label{eq:pness}
\end{equation}
 with
\begin{eqnarray}
 & A=\frac{\rho c_{p}(\gamma-1)}{2T}\left[1-\frac{1}{\alpha c_{p}}\left(\frac{\partial c_{p}}{\partial T}\right)_{P}+\frac{1}{\alpha^{2}}\left(\frac{\partial\alpha}{\partial T}\right)_{P}\right]\nonumber \\
\label{eq:bigA}
\end{eqnarray}
Here $c_{p},\gamma,\alpha$ are, respectively the specific heat at
constant pressure, the ratio of specific heats, and the coefficient
of thermal expansion. More generally, the long distance part of the
fluctuating pressure is \cite{Kirkpatrick_DeZarate_Sengers_2013,Kirkpatrick_DeZarate_Sengers_2014}
\begin{equation}
\tilde{P}(\mathbf{x})=A[\delta T(\mathbf{x})]^{2}\label{eq:flucpress}
\end{equation}
Averaging this over the NESS and using Eq.(\ref{eq:delTk}) gives
Eq.(\ref{eq:pness}).

The pressure fluctuations are $\delta\tilde{P}(\mathbf{x})=A[(\delta T(\mathbf{x}))^{2}-\langle(\delta T(\mathbf{x}))^{2}\rangle_{NESS}]$
and the non-equilibrium correlation function is $C_{PP}(\mathbf{x},\mathbf{y})=<\delta\tilde{P}(\mathbf{x})\delta\tilde{P}(\mathbf{y})>_{NESS}$.
Using a Gaussian approximation, we find this correlation function to be,
\begin{equation}
C_{PP}(\mathbf{x},\mathbf{y})=2A^{2}[<\delta T(\mathbf{x})\delta T(\mathbf{y})>_{NESS}]^{2}
\end{equation}
The growth of the temperature correlations with distance implies that
the spatially averaged $C_{PP}$ behaves as,
\begin{equation}
\bar{\Delta}_{P,NESS}\propto[\frac{\ell A(\Delta T)^{2}}{L}]^{2}\propto\frac{1}{L^{2}}\label{eq:mnsqpress}
\end{equation}
while in equilibrium,
\begin{equation}
\bar{\Delta}_{P,eq}=\frac{nk_{B}T}{L^{3}}\left(\frac{\partial P}{\partial n}\right)_{s}\propto\frac{1}{L^{3}}\label{eq:mnsqpresseq}
\end{equation}
Again, the non-equilibrium fluctuations are large compared to the
equilibrium ones at large length scales. Comparing Eqs.(\ref{eq:mnsqpress}) and (\ref{eq:mnsqpresseq}) and using Eq.(\ref{eq:Lovrsig}), this roughly occurs when $L/\sigma > \epsilon^4$, or, in typical experiments, $L/\sigma > 10^4$.

If we assume Gaussian behavior then the normal stress fluctuation
distributions in the NESS and in equilibrium are

\begin{equation}
\mathcal{P}_{NESS}[\delta P]\sim\exp[-\frac{L^{2}}{2}\left(\frac{\delta P}{A\ell\left(\Delta T\right)^{2}}\right)^{2}]\label{eq:pressdist}
\end{equation}
and 
\begin{equation}
{\cal P}_{eq}(\delta P)\sim\exp\left[-\frac{L^{3}}{2nk_{B}T}\left(\frac{\partial n}{\partial P}\right)_{s}\left(\delta P\right)^{2}\right].\label{eq:deltaPeq}
\end{equation}
That is, in three-dimensional non-equilibrium systems the normal stress
fluctuations scale as $\delta P\sim1/L\sim1/N^{1/3}$ (compared to
$1/N^{1/2}$ in equilibrium systems).

As a final example we consider the NESS fluctuations in the heat current
in some direction, $\delta j_{H,\alpha}$, where $\alpha=(x,y,z)$. There are both linear and nonlinear (analogous to Eq.(\ref{eq:flucpress})) contributions to the heat flux. The linear contribution can be obtained by identifying the microscopic fluctuating heat current as $\mathbf{\delta j}_{H}(\mathbf{x})=-\lambda\mathbf{\nabla}\delta T(\mathbf{x})$
with $\delta T$ the fluctuating temperature gives,
\begin{equation}
C^L_{H,\alpha\beta}(\mathbf{x},\mathbf{y})=\lambda^{2}\nabla_{x,\alpha}\nabla_{y,\beta}<\delta T(\mathbf{x})\delta T(\mathbf{y})>_{NESS}\label{eq:temperature fluctuation}
\end{equation}
The long ranged part of the temperature fluctuation in Eq.(\ref{eq:temperature fluctuation})
is given by Eq.(\ref{eq:delTk}), with the final result for $C^L_{H}$
in Fourier space given by,
\begin{equation}
C^L_{H,\alpha\beta}(\mathbf{k})=\frac{\ c_{p}k_{B}T\lambda}{(\nu+{D}_{T})\mathrm{k}^{\mathrm{4}}}\hat{k}_{\alpha}\hat{k}_{\beta}[k_{\parallel}\nabla T]^{2}\label{eq:Chtcurr}
\end{equation}
This result can also be directly obtained with kinetic theory methods \cite{Kirkpatrick_Cohen_Dorfman_1982B}.
Here the $\hat{\mathbf{k}}$ denotes unit vectors. The $1/k^{2}$
dependence in Eq.(\ref{eq:Chtcurr}) implies that for systems in $d$
dimensions, $C^L_{H}(r\rightarrow\infty)\sim1/r^{d-2}$, that is, power
law correlations. The angular factors don't really change this in
general. For example, it is easy to see that if in three-dimensions
one look in the middle of the fluid at $x_{z}=y_{z}$ and considers
correlations along that plane (perpendicular to the direction of the
temperature gradient) as a function of $r_{\bot}$ one finds that
the $z-$component of the current fluctuations is $\sim1/r_{\bot}$.
Physically this means that if there is a current fluctuation in the
$\mathit{wrong}$ direction then the fluctuations in that entire plane
will likely also be in the $\mathit{wrong}$ direction. To obtain
a better measure of these correlations we consider the spatial average
of $C^L_{H}(\mathbf{x})$,
\begin{equation}
\bar{\Delta}^L_{H,NESS}\propto\frac{k_{B}Tc_{p}\lambda}{(D_{T}+\nu)}L^{2-d}(\nabla T)^{2}\propto\frac{(\Delta T)^{2}}{L^{d}}\label{eq:spatavheatcurr}
\end{equation}
That is, even though the current correlations are of long-range, they
are weighted by a factor of $(\nabla T)^{2}\propto1/L^{2}$ , for
fixed $\Delta T$, so that qualitatively scales just like $\bar{\Delta}_{H,eq}\propto1/L^{d}$
and is, in fact, small compared to $\bar{\Delta}_{H,eq}$ because
it is of relative order $\left(\frac{\Delta T}{T}\right)^{2}$. 

In analogy with Eq.(\ref{eq:flucpress}), the non-linear portion of the fluctuating heat current is $\mathbf{j}_{H}(\mathbf{x})\propto{\delta}T(\mathbf{x})\mathbf{u}(\mathbf{x})$,
with $\mathbf{u}$ the fluctuating fluid velocity. The fluctuation correlation of this quantity gives $C^{NL}_{H}(\mathbf{x})$. It can be readily computed \cite{Kirkpatrick_Cohen_Dorfman_1982A,DeZarate_Sengers_2006} and
in space it decays as $\sim(\nabla T)^{2}/r^{2d-4}\sim(\Delta T)^{2}/[L^{2}r^{2d-4}]$. To obtain a better measure of these correlations we consider the spatial average of $C^{NL}_{H}(\mathbf{x})$,
\begin{equation}
\bar{\Delta}^{NL}_{H,NESS}\propto\frac{(\Delta T)^{2}}{L^{2d-2}}\label{eq:spatavNLheatcurr}
\end{equation}
These correlations are even weaker then those from the linear portion of the fluctuating heat current and can therefore be neglected.

We conclude with a number of remarks: 
\begin{enumerate}
\item The result for $\bar{\Delta}_{P,NESS}$, Eq.(\ref{eq:mnsqpress}),
has an important implication. In Landau and Lifshitz (L\&L) fluctuating
hydrodynamics an input is that the fluctuating stress tensor is local
in space and time. Averaging the normal component of the L\&L fluctuating
stress tensor correlation function in three-dimensions over space
and and a microscopic time interval leads to a quantity $\bar{\Delta}$
that scales like $\bar{\Delta}\propto1/L^{3}$ which should be compared
with $\bar{\Delta}_{P,NESS}\propto1/L^{2}$. That is, our results
indicate that there are important
corrections to the usual linearized L\&L equations due to the very
long-range correlations that exist, in at least some NESS. 

\qquad{}In a non-linear fluctuating hydrodynamic description these
anomalous terms will be generated by fluctuation or renormalization
effects.
\item Compared to the pressure or stress fluctuations both the linear and non-linear heat current fluctuations are of relatively short range. This result is actually important: It is consistent with the local assumption for the fluctuating heat current
in the L\&L approach. This in part explains why conventional linearized
fluctuating hydrodynamics can be used to successfully compute the very long range
NESS temperature fluctuations \cite{Ronis_Procaccia_1982,DeZarate_Sengers_2006}.
\item As another measure of just how anomalous the normal stress fluctuations
are, note that if a fluctuation as large as the average value of $P_{NESS}$
is considered, Eq.(\ref{eq:pness}), in $\mathcal{P}_{NESS}[\delta P]$
then the probability of that fluctuation is of $O(L^{0})$. That is,
the normal stress is not a self-averaging quantity. 
\item The correlations in a fluid with a constant velocity gradient (that
is, the case studied by ECM) are very different (see, for example,
\cite{Lutsko_Dufty_2002}) then either a fluid with a temperature
gradient, or in a mixture with a concentration gradient. Although
there are power law correlations in the velocity gradient case, they
are sufficiently weak that the normalized NESS stress-tensor correlation
function does scale like $\sim1/N^{1/2}$. This is true whether the
velocity gradient, $\nabla u$, with $u$ the hydrodynamic velocity,
is fixed or if $\nabla u=\Delta u/L$ is used, and $\Delta u$ is
fixed.
\item In our presentation we have ignored a discussion of the boundaries,
that must be present in this NESS. For the case of perfectly conducting
walls the temperature fluctuation problem in a bounded geometry has
been discussed and solved elsewhere \cite{DeZarate_Sengers_2006}. 
\item Structurally identical results as those given here for a single component
fluid with a temperature gradient are obtained in a fluid mixture
with either a concentration gradient or a temperature gradient(see \cite{DeZarate_Sengers_2006}
and references therein). In this case, the temperature fluctuations,
the concentration fluctuations, and the normal stress, or pressure,
fluctuations are all found to obey anomalous statistics. 
\item The profound difference between correlations in non-equilibrium and
equilibrium systems means that many of the theoretical techniques
developed to describe equilibrium systems cannot be applied to non-equilibrium
systems. Non-equilibrium quantities do
not have virial expansions \cite{Dorfman_Cohen_1965}. A local expansion of the fluxes or currents
in terms of powers of the gradients is also not possible \cite{Dorfman_Kirkpatrick_Sengers_1994,Belitz_Kirkpatrick_Vojta_2005,Ernst_et_al_1978}.
Other techniques such as maximizing an entropy (for example, the so-called
max cal method \cite{Presse_et_al_2013}) to obtain a non-equilibrium
distribution function may not work, at least in
their most naive form. 
\end{enumerate}
This work has been supported by the National Science Foundation under
grant number DMR-1401449.

\end{document}